\documentclass[twocolumn]{article}
\usepackage[dvips]{graphics,color}

\begin{document}                  

\renewcommand{\thefootnote}{\fnsymbol{footnote}}

\title{On the crystal structure of lithium peroxide, Li$_2$O$_2$\footnotemark[1]}

\author{Luis Guillermo Cota\\
Pablo de la Mora\footnotemark[2]\\
Departamento de F\'{\i}sica, Facultad de Ciencias.\\ 
Universidad Nacional Aut\'onoma de M\'exico.\\ Circuito Exterior, 
Ciudad Universitaria, M\'exico, D.F. 04510, M\'exico}

\maketitle

\begin{abstract}
The two published lithium peroxide structures, both ascribed to the hexagonal 
$P \overline{6}$ space group, were subjected to reinterpretation and 
another more symmetric structure, now belonging to the $P 6_3/m m c$ space group, was found. Detailed density-functional quantum mechanical calculations and crystal structure optimisations were carried out on both structures and the energetic arguments obtained therewith helped to rule out one of them.
\end{abstract}

\footnotetext[1]{{\protect Accepted for publication in {\em Acta Crystallographica}, Section B.}}
\footnotetext[2]{{\protect Corresponding author.
Electronic address: \\delamora@servidor.unam.mx}} 

\section{Introduction}

Despite its use as an oxidizing agent, as a regenerator for confined breathable atmospheres and as a fuel cell material, lithium peroxide has seemingly not been sufficiently studied, to the point that even its crystalline structure has not been unambiguously determined. 
	
Crystallographic databases give two different accounts of the lithium peroxide structure, the first one proposed by F\'eher {\em et\,al.} (1953), and the other one by F\"oppl (1957). Whether these two structures represent isomorphs is not substantiated by experimental evidence other than the interpretation of X-ray diffraction data, which is not entirely suitable for lithium compounds due to the fact that lithium is a poor X-ray scatterer. Therefore, since the issue of deciding for either one of the proposed structures is not completely resolved, further argumentation is in order.

In the course of a density functional, quantum mechanical study of lithium peroxide, we found out that both proposed structures could be better accounted for by another more symmetric hexagonal space group. On the other hand, optimisation of atomic coordinates and cell parameters shows that the F\"oppl structure is substantially more favourable in terms of energy and in terms of its ability to reproduce the experimental X-ray spectrum of the peroxide.

This study relies on the fact that modern high-quality DFT calculations are nowadays very much able to make not only qualitative but also accurate quantitative predictions.

\section{Calculation details}

The calculations were carried out using the {\em WIEN2k} code (Blaha {\em et\,al.}, 2001), which is a full potential-linearised augmented plane wave (FP-LAPW) method based on the density functional theory (DFT). The generalised gradient approximation (Perdew {\em et\,al.}, 1996) was used for the treatment of the exchange-correlation interactions. The number of {\em k}-points used was 1000 (72 in the irreducible wedge). 
For the number of plane waves the applied criterion was $R_{MT}^{max} \times
K^{max} = 8$ (muffin-tin radius and maximum number of plane waves,
respectively). The same $R_{MT}$ was used for each atom in all the crystal structures, $R_{MT}({\rm Li}) = 1.8$ a.u., $R_{MT}({\rm O}) = 1.4$ a.u., except for the original F\'eher {\em et\,al.} structure, where $R_{MT}({\rm O}) = 1.2$ a.u., was used.
 The space group analysis was done with the program {\em sgroup} (Yanchitsky \& Timoshevskii, 2001) contained in the {\em WIEN2k} package. For the crystal structure visualisation the {\em XCrySDen} package (Kokalj, 1999) was used.

\section{Discussion}

\subsection{Crystal structure simplification}

The crystal structure of lithium peroxide was reported some 50 years ago by F\'eher {\em et\,al.}, (from now on $Str1$), in 1953, and by F\"oppl, (from now on $Str2$), in 1957. Both authors proposed their respective structural models based on the good agreement with their respective X-ray data. F\"oppl, however, based on a systematic study of several alkali peroxides, claimed that his structural assignment was more in accordance with the general peroxide construction principles, although he provided no further evidence pertaining to lithium peroxide itself.

$Str1$ is presented in Table 1.
From the atomic positions it can be readily seen that, by shifting the three Li\,3 atoms by half cell in the {\em a} or {\em b} direction, they fall into the Li\,1 positions, and the same is valid for the other atoms, such that Li\,4 becomes Li\,2, O\,3 becomes O\,1 and O\,4 becomes O\,2. This clearly suggests that the unit cell can be reduced to $a'$ = $a/2$ and $b'$ = $b/2$ ($a'$ = $b'$ = $3.1525$\,\AA), the 2 g (0 0 {\em z}) positions remain the same, and the 2 i ($1/3$  -$1/3$ {\em z}) positions change to 2 d (-$1/3$ $1/3$ {\em z}). The reduced structure is therefore as in Table 2.

This simpler cell, with only 4 lithiums and 4 oxygens has no inversion symmetry. However, by shifting all atoms by $(1/3$ -$1/3$ $1/4)$  Li\,1 and Li\,2 are now connected by inversion, as well as O\,1 and O\,2. The new cell belongs to a different space group, as is shown in Table 3.

Lithium atoms are arranged in the wurtzite structure, which is similar to the $h.c.p.$ (hexagonal close packed) structure, but instead of having alternating triangular layers, $BCBC...$, the layers are repeated twice, $BBCCBBCC...$. At the same time, oxygen atoms occupy alternate sites, that is, $CCBBCCBB...$; therefore, the new structure can be viewed as two interpenetrating wurtzite structures, one shifted from the other by $1/2$ cell in the {\em c} direction. This is indeed an open structure with the $A$ sites vacant. In this structure the vertical distances between two lithiums and between two oxygens cannot be determined by symmetry (these distances, which will be referred to as $2\alpha$, are $2\alpha({\rm Li}) = 0.316$ ($2.43$\,\AA), and $2\alpha({\rm O}) = 0.166$ ($1.28$\,\AA)).

$Str2$ is presented in Table 4. Again, inversion symmetry can be incorporated by shifting the atoms by $(1/3$ -$1/3$ $1/4)$. Now Li\,1 and Li\,2 are connected by inversion and both therefore become Li\,2, Li\,3 becomes Li\,1, and finally also O\,1 and O\,2 are connected by inversion. The new cell is shown in Table 5. As can be seen, this structure is quite similar to the reduced  $Str1$ cell (Table 3), where oxygen ions have the same 4\,f position, but with a larger separation, $2\alpha({\rm O}) = 0.204$ ($1.56$\,\AA). Lithium atoms, however, have different positions: instead of being on the oxygen planes they are {\em between} the planes. The arrangement can be described as $AcBcAbCbA$, where capital letters refer to lithium and small letters to oxygen. This can be seen as a close packed arrangement, compressed in the $c$ direction. The sequence $BcAbC$ forms a cubic close packed arrangement {\em (ccp)}, followed by the reverse sequence, $CbAcB$, which also forms a {\em ccp} arrangment. Therefore, $Str2$ can be regarded as mirrored layers of {\em ccp} arrangements. Lithium atoms by themselves form a close packed structure, $ABACA$, which is still compressed in the $c$-direction, with $c/a = 2.43$, compared to the ideal hard sphere packing of 3.26.

\subsection{Crystal structure optimisation}

To elucidate whether these two crystal structures represent isomorphs of $Li_2O_2$, electronic structure calculations were performed on the symmetry-optimised version of each peroxide structure. Both structures were found to be strained ({\em i.e.,} the calculations showed that there were non-negligible net forces acting on the atoms), therefore, a full optimisation of the atomic coordinates and cell parameters was granted.

\subsubsection{F\'eher {\em et\,al.} structure, $Str1$.}

This structure showed large forces on its atoms, with $F(Li)$\,=\,$-10$\,mRy/a.u. and $F(O)$\,=\,$-374$\,mRy/a.u, (Ry: Rydberg = 13.6 eV) according to our calculations. Relaxing the atomic coordinates (until the forces became negligible, that is, less than $1$\,mRy/a.u.) produced drastic changes (see Table 6): there is a total energy reduction of $-0.11$\,Ry, whereas the O--O bond lenght increases from 1.28\,\AA\, to $1.50$\,\AA.

On his part, F\"oppl argued that such a short 1.28\,\AA\, {\protect O--O} bond in the F\'eher {\em et\,al.} structure would reproduce the X-ray diffraction pattern well. However, in analogy with his homologous peroxide determinations,  F\"oppl concluded that this O--O bond should be of the order of 1.5\,\AA, and with such large bond the X-ray diffraction pattern would be discernibly different. Therefore, according to  F\"oppl,  $Str1$ must be wrong.

Full optimisation of this structure was carried out in sucessive cycles of force,  cell volume and $c/a$ optimisation steps. In the last cycle the changes in cell volume and $c/a$ were 0.39\,\% and 0.29\,\%, respectively. The final forces were $F(Li)=1.3$\,mRy/a.u. and $F(O)=2.8$\,mRy/a.u. The optimisation after the relaxation of the atomic coordinates did not have such a large impact on the energy or on the O--O bond lenght, but produced a considerable $15$\,\% cell volume increase and a $2.7$\,\%  reduction in the $c/a$ ratio. Such large discrepancies with the originally reported experimental values must necessarily entail a misinterpretation of the X-ray diffraction measurements, which should perhaps have been evident even in the $1953$ X-ray diffraction patterns. 

\subsubsection{F\"oppl structure, $Str2$.}

Our calculations show that there is a moderate force on the oxygen atom, of $11$\,mRy/a.u., in this structure. Full optimisation of the structure (see Table 7) produces a moderate volume increase of $3.6$\,\% and some other quite small changes: $-1.8$\,mRy total energy reduction, an {\protect O--O} bond length reduced from 1.56\,\AA\, to 1.55\,\AA\, and a $c/a$ ratio change of 0.31\,\%.
The changes in volume and $c/a$ were 0.22\% and 0.15\%, respectively, in the last cycle of optimisation. The final force was $F(O)$\,=\,$1.6$\,mRy/a.u.

There is a large initial energy difference between $Str2$ and $Str1$ of $-110$\, mRy/peroxide unit, which reduces after optimisation to only  $-39$\,mRy/peroxide. However, this energy difference is still large enough (Fournier, 1993), which constitutes further evidence that lithium peroxide does not have the $Str1$ structure.

Therefore, our results show that lithium peroxide has the same crystal structure proposed by F\"oppl, but now with inversion symmetry (space group $194$)  and slightly modified atomic positions (Figure\,1 and Table 5). The optimized structure has the following cell parameters: $a$\,=\,$3.1830$\,\AA\, and $c$\,=\,$7.7258$\,\AA, with $\alpha$\,=\,$1.003$ (instead of $\alpha$\,=\,$1.02$).

\section{Conclusions}

Two crystal structures have been proposed for lithium peroxide. These structures
 were formerly ascribed to the space group $P \overline{6}$ (SG \#174). It was 
found that they could be better described by the $P 6_3/m m c$ space group (SG \#194).

The results presented in this work, from calculations performed with high-quality DFT codes, offer a strong argument against the viability of the structure proposed by F\'eher {\em et\,al.} in 1953. They do, however, confirm the correctness of the structure proposed by F\"oppl. Furthermore, on these grounds, it was possible to present improved values for the Li$_2$O$_2$ cell parameters.\\

The authors wish to thank the valuable insight provided by professors Miguel Castro and Gustavo Taviz\'on, of the Facultad de Qu\'{\i}mica, UNAM.\\

\section{References}

Blaha, P., Schwarz, K., Madsen, G. K. H., Kvasnicka, D. \& Luitz, J. \emph{WIEN2k}, an augmented plane wave + local orbitals program for calculating crystal properties (Karlheinz Schwarz, Techn. Universit\"at Wien, Austria) (2001). ISBN 3-9501031-1-2.
\\
\noindent F\'eher, F., von Wilucki, I. \& Dost G. (1953). \emph{Chemische Berichte} \textbf{86}, 1429-1437.
\\ 
\noindent F\"oppl, H. (1957). \emph{Zeitschrift f\"ur anorganische und allgemeine Chemie} \textbf{291}, 12-50.
\\ 
\noindent Fournier, R. (1993). \emph{Journal of Chemical Physics} \textbf{99}, 1801-1815.
\\ 
\noindent Kokalj, A. (1999). \emph{Journal of Molecular Graphics and Modelling} \textbf{17}, 176-179.
\\ 
\noindent Perdew, J. P., Burke, S., \& Ernzerhof, M. (1996). \emph{Physical Review Letters} \textbf{77}, 3865-3868.
\\ 
\noindent Yanchitsky, B. \& Timoshevskii T. (2001). \emph{Computer Physics Communications} \textbf{139}, 235-242.

\begin{table}
\begin{tabular}{|c|c|r|r|r|r|}
\hline
Atom, \# & site & x & y & z \\
\hline
\hline
Li\,1	& 2 g	&  0	&  0  &      $\pm0.158$ \\
\hline
Li\,2	&2 i     & -$1/3$&	$1/3$&	$1/2\pm0.158$ \\
\hline
Li\,3	&6 l	&$1/2$&	$1/2$ & $\pm0.158$\\
	&   	&0	&$1/2$&     $\pm0.158$\\
	&  	&$1/2$	&0       & $\pm0.158$\\
\hline
Li\,4	&6 l	&$1/6$  &    -$1/6$	&$1/2\pm0.158$\\
	&          &-$1/3$   &   -$1/6$	&$1/2\pm0.158$\\
	& &$1/6$&	$1/3$&	$1/2\pm0.158$\\
\hline
O\,1	&2 g 	&0&	0&	$1/2\pm0.083$\\
\hline
O\,2	&2 i      & -$1/3$    &   $1/3$    & $\pm0.083$\\
\hline
O\,3	&6 l 	&$1/2$	&$1/2$&	$1/2\pm0.083$\\
	&   	&0	&$1/2$	& 	$1/2\pm0.083$\\
	&   	&$1/2$	&0	&$1/2\pm0.083$\\
\hline
O\,4	&6 l	&$1/6$  &    -$1/6$  &   $\pm0.083$\\
	&        &  -$1/3$  &    -$1/6$ &   $\pm0.083$\\
	&   	&$1/6$&	$1/3$&     $\pm0.083$\\
\hline
\end{tabular}
\\
\caption{
Structure 1, original (F\'eher {\em et\,al.}, 1953).
\newline
Space group: $P \overline{6}$ 
(SG \#174), hexagonal.
\newline
Unit Cell: 6.305 6.305 7.71 90. 90. 120.
}
\end{table}

\begin{table}
\begin{tabular}{|c|c|r|r|r|r|}
\hline
Atom, \# & site & x & y & z \\
\hline
\hline
Li\,1	&2 g	&0	&0   &     $\pm0.158$\\ 
\hline
Li\,2	&2 d&	1/3  &    -$1/3$&	$1/2\pm0.158$ \\
\hline
O\,1	&2 g &	0&	0&	$1/2\pm0.083$\\
\hline
O\,2	&2 d&	$1/3$  &    -$1/3$  &   $\pm0.083$\\
\hline
\end{tabular}
\caption{Structure 1, reduced.
\newline
Space group: $P \overline{6}$ (SG \#174), hexagonal.
\newline
Unit Cell: 3.1525 3.1525 7.71 90. 90. 120.}
\end{table}

\begin{table}
\begin{tabular}{|c|c|r|r|r|r|}
\hline
Atom, \# & site & x & y & z \\
\hline
\hline
Li\,1&	4 f&	$1/3$  &    -$1/3$  &    $1/4\pm0.158$ \\
 	  &         &    -$1/3$&	$1/3$  &   -$1/4\pm0.158$ \\
\hline
O\,1&	4 f   &     $1/3$   &  -$1/3$  &   -$1/4\pm0.083$\\
      &          &   -$1/3$&	$1/3$  &    $1/4\pm0.083$\\
\hline
\end{tabular}
\caption{Structure 1, symmetrised.
\newline
Space group: $P 6_3/m m c$ (SG \#194), hexagonal.
\newline
Unit Cell: 3.1525 3.1525 7.71 90. 90. 120.}
\end{table}
 
\begin{table}
\begin{tabular}{|c|c|r|r|r|r|}
\hline
Atom, \# & site & x & y & z \\
\hline
\hline
Li\,1	&1 a&	0	&0	&0\\
\hline
Li\,2	&1 d&	$1/3$  &    -$1/3$	&1/2\\
\hline
Li\,3	&2 i   &    -$1/3$&	$1/3$   &  $\pm1/4$\\
\hline
O\,1	&2 g&	0&	0&	$1/2\pm0.102$\\
\hline
O\,2	&2 h&	$1/3$&      -$1/3$&     $\pm0.102$\\
\hline
\end{tabular}
\caption{Structure 2, original (F\"oppl, 1957).
\newline
Space group: $P \overline{6}$ (SG \#174), hexagonal.
\newline
Unit Cell: 3.142 3.142 7.65 90. 90. 120.}
\end{table}

\begin{table}
\begin{tabular}{|c|c|r|r|r|r|}
\hline
Atom, \# & site & x & y & z \\
\hline
\hline
Li\,1	&2 a	&0	&0&	0\\
	&     &	0&	0&	$1/2$\\
\hline
Li\,2	&2 c&	$1/3$   &   -$1/3$&	$1/4$\\
 	&     &       -$1/3$&	$1/3$ &     -$1/4$\\
\hline
O\,1&	4 f&	$1/3$ &     -$1/3$ &     -$1/4\pm0.102$\\
	&       &      -$1/3$	&$1/3$	&$1/4\pm0.102$\\
\hline
\end{tabular}
\caption{Structure 2, symmetrised.
\newline
Space group: $P 6_3/m m c$ (SG \#194), hexagonal.
\newline
Unit Cell: 3.142 3.142 7.65 90. 90. 120.}
\end{table}
 
\begin{table}
\begin{tabular}{|c|c|c|c|c|c|c|}
\hline
Structure  & $a$ & $c$ & 2$\alpha(Li)$ & 2$\alpha(O)$ & O--O & Energy\\
\hline
\hline
Unrelaxed & 3.1525  &  7.71   &  0.316   & 0.166 &   1.28   &   -662.2907\\
\hline
Force relaxed & 3.1525  &  7.71   &  0.2952   & 0.1943 &   1.50   &   -662.4002\\
\hline
Fully relaxed & 3.3339  &  7.9300   &  0.2937   & 0.1923 &   1.53   &   -662.4262\\
\hline
\end{tabular}
\caption{Optimisation of structure 1.
\newline
(Forces are in mRy/a.u., energies in Ry and bond distances in \,\AA)}
\end{table}

\begin{table}
\begin{tabular}{|c|c|c|c|c|c|c|}
\hline
Structure  & $a$ & $c$ & 2$\alpha(O)$ & O--O & Energy\\
\hline
\hline
Unrelaxed & 3.142  &  7.65  & 0.204 &   1.5606   &   -662.5025\\
\hline
Fully relaxed & 3.1830  &  7.7258 & 0.2006 &   1.5500   &   -662.5043\\
\hline
\end{tabular}
\caption{Optimisation of structure 2.
\newline
(Forces are in mRy/a.u., energies in Ry and bond distances in \,\AA)}
\end{table}

\begin{figure}
\resizebox{8.4cm}{!}{\includegraphics{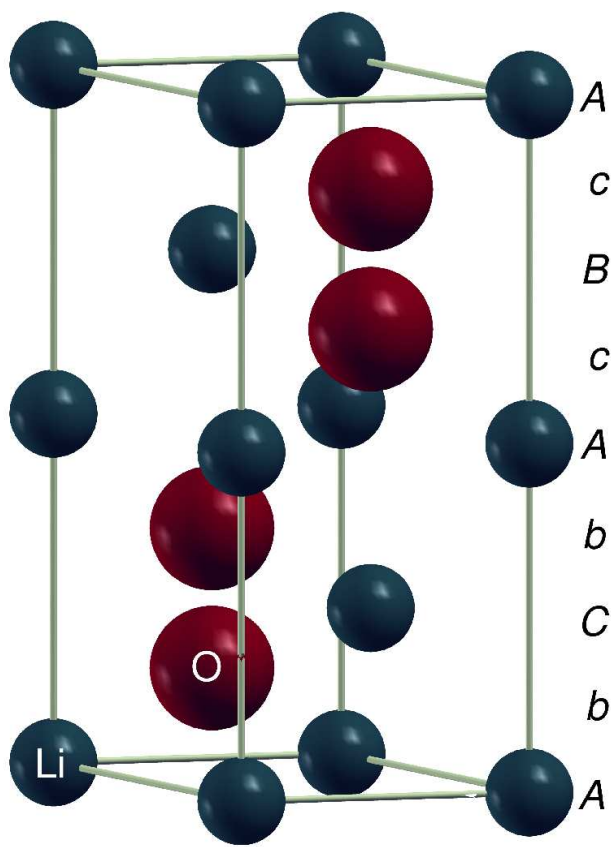}}
\caption{Optimised Li$_2$O$_2$ structure.}
\end{figure}

\end{document}